\documentclass[twocolumn]{revtex4}
\usepackage{graphicx,epsf}
\usepackage{color}
\usepackage{dcolumn}
\usepackage{amsfonts}
\usepackage{mathrsfs}

\usepackage{subfig}

\begin{document}

\title{Effects of energetic particles on    zonal flow generation by toroidal Alfv\'en eigenmode}

\author{Z. Qiu$^{1}$, L. Chen$^{1, 2}$ and F. Zonca$^{3, 1}$}

\affiliation{$^1$Institute for    Fusion Theory and Simulation and Department of Physics, Zhejiang University, Hangzhou, P.R.C\\
$^2$Department of   Physics and Astronomy,  University of California, Irvine CA 92697-4575, U.S.A.\\
$^3$ ENEA C. R. Frascati, C. P.
65-00044 Frascati, Italy}

\begin{abstract}
Generation of zonal flow (ZF) by energetic particle (EP) driven toroidal Alfv\'en eigenmode (TAE) is investigated using nonlinear gyrokinetic theory. It is found that, nonlinear resonant EP contribution  dominates over the usual Reynolds and Maxwell stresses due to thermal plasma nonlinear response. ZF can be forced driven  in the linear growth stage of TAE, with the growth rate being twice the TAE   growth rate. The ZF generation mechanism is shown to be related to polarization induced by resonant EP  nonlinearity. The generated ZF has both  the usual meso-scale and   micro-scale radial structures. Possible consequences of this forced driven ZF on the nonlinear dynamics of TAE are also discussed.
\end{abstract}

\maketitle

Understanding the nonlinear dynamics of shear Alfv\'en waves (SAW) is of crucial importance to future burning plasmas with  energetic particle (EP) population such as fusion-$\alpha$s significantly contributing to the overall plasma energy density \cite{LChenRMP2016}.  With  frequency comparable to the characteristic frequencies of EPs, and group velocities mainly along magnetic field lines, SAWs are expected to be driven unstable by resonant EPs \cite{YKolesnichenkoVAE1967,AMikhailovskiiSPJ1975,MRosenbluthPRL1975,LChenPoP1994}; leading to  EP transport and degradation of overall confinement, as reviewed in Ref. \citenum{LChenRMP2016}. Toroidal Alfv\'en eigenmode (TAE) \cite{CZChengAP1985,GFuPoFB1989}, excited inside the toroidicity-induced SAW continuum gap to minimize continuum damping, is one of most dangerous candidates for effectively scattering EPs.

There are two routes for the nonlinear saturation of TAEs, i.e., nonlinear wave-particle and nonlinear wave-wave interactions \cite{LChenPoP2013}.
Wave-particle phase space nonlinearity \cite{FZoncaNJP2015}, e.g., wave-particle trapping,  describes the nonlinear distortion of the   EP distribution function; and leads to SAW saturation as the wave-particle trapping frequency, proportional to square root of the mode amplitude, is comparable with linear growth rate \cite{TOneilPoF1965,HBerkPoFB1990a,HBerkPoFB1990c,JZhuNF2014}. On the other hand,  wave-wave coupling accounts for the transfer of TAE wave energy away from the most unstable modes. Among various wave-wave nonlinearities, generation of zonal structures (ZS)  is of particular importance.
Chen et al \cite{LChenPRL2012} investigated the nonlinear excitation of zero frequency zonal structure (ZFZS) by   TAE with a prescribed amplitude, and found that finite amplitude TAE can excite ZFZS via modulational instability at a rate proportional to the amplitude of the pump TAE. Meanwhile, zonal current with lower excitation threshold could be preferentially excited in specific plasma equilibria, which, however, do not reflect typical experimental tokamak plasmas \cite{LChenPRL2012}. Numerical simulations of nonlinear dynamics of EP driven TAE are carried out by both hybrid code \cite{YTodoNF2010} and PIC code \cite{ZWang2016}, and found that zonal flow (ZF) is excited by forced driven process, with the ZF growth rate being twice of TAE  growth rate.  In this paper, we will clarify the ``discrepancies" between analytical theory and simulation, with emphasis on the important role played by EPs \cite{FZoncaVarenna2000,LChenRMP2016}. Our results indicate that  there is no conflict  between   analytical theory \cite{LChenPRL2012} and numerical simulations \cite{YTodoNF2010,ZWang2016}; in fact, they describe two nonlinear processes that occur at different stages of the TAE nonlinear dynamics.

To derive the fully nonlinear equations describing nonlinear ZFZS generation by TAE, we take $\delta\phi$ and $\delta A_{\parallel}$ as the field variables. Here, $\delta\phi$ and $\delta A_{\parallel}$ are  the scalar potential and parallel component of vector potential to the equilibrium magnetic field, respectively. An alternative field variable $\delta\psi\equiv \omega\delta A_{\parallel}/(ck_{\parallel})$ is also adopt here, and one has $\delta\psi=\delta\phi$ in the ideal MHD limit.  For the nonlinear interactions between TAE and ZFZS, we take $\delta\phi=\delta\phi_Z+\delta\phi_T$, with $\delta\phi_T=\delta\phi_0+\delta\phi_{0}^*$. We assume the  well-known ballooning-mode decomposition in the $(r,\theta,\phi)$ field-aligned toroidal flux coordinates:
\begin{eqnarray}
\delta\phi_0=A_0e^{i(n\phi-m_0\theta-\omega_0t)}\sum_j e^{-ij\theta}\Phi_0(x-j).\nonumber
\end{eqnarray}
Here, $n$ is the toroidal mode number,  $(m=m_0+j)$ is the poloidal mode number with $m_0$ being its reference value satisfying $nq(r_0)=m_0$, $q(r)$ is the safety factor, $x=nq-m_0\simeq nq'(r-r_0)$, $\Phi_0$ is the fine scale structure associated with $k_{\parallel}$ radial dependence and magnetic shear, and  $A_0$ is the radial envelope
\begin{eqnarray}
A_0=\hat{A}_0e^{i\int \hat{k}_{0,r} dr}\nonumber
\end{eqnarray}
with $\hat{A}_0$ being the envelope amplitude and $\hat{k}_{0,r}\equiv nq'\theta_k$ being the radial envelope wavenumber in the ballooning  representation. For ZFZS, we take
\begin{eqnarray}
\delta\phi_Z=A_Ze^{-i\omega_Z t}\sum_m\Phi_Z \nonumber
\end{eqnarray}
with $\Phi_Z$ being the fine radial structure \cite{ZQiuNF2016}, and $A_Z$ being the usual ``meso"-scale structure
\begin{eqnarray}
A_Z=\hat{A}_Ze^{i\int \hat{k}_Zdr}.\nonumber
\end{eqnarray}

The nonlinear equation for ZF can be derived from nonlinear vorticity equation
\begin{eqnarray}
&&(e^2/T_i)\langle(1-J^2_k)F_0\rangle\delta\phi_Z-\sum_s\left\langle(e_s/\omega)J_k\omega_d\delta H\right\rangle_Z\nonumber\\
&=&-ic\Lambda_Z\left[ c^2k''^2_{\perp}\partial_l\delta\psi_{k'}\partial_l\delta\psi_{k''}/(4\pi\omega_{k'}\omega_{k''})\right.\nonumber\\
&&\left.+\left\langle e(J_kJ_{k'}-J_{k''})\delta L_{k'}\delta H_{k''}\right\rangle\right]/(\omega_ZB_0),
\label{vorticityequation}
\end{eqnarray}
where the two explicitly nonlinear terms on the right hand side are, respectively,  Maxwell and Reynolds stresses, the subscripts $s=i, e, E$ denotes particle species, and
\begin{eqnarray}
\Lambda_k\equiv \sum_{\mathbf{k}'+\mathbf{k}''=\mathbf{k}}\hat{\mathbf{b}}\cdot\mathbf{k}''\times\mathbf{k}'.\nonumber
\end{eqnarray}
Here, $\mathbf{k}$ are defined as the operators  for spatial derivatives, and we have
\begin{eqnarray}
\mathbf{k} \delta\phi &\equiv&[k_{\parallel}\mathbf{b}+ k_{\theta}\hat{\mathbf{\theta}}+\left(\hat{k}_r-inq'\partial_x \ln\Phi\right)\hat{\mathbf{r}}]\delta\phi.\nonumber
\end{eqnarray}
We note that  EPs,  with $|k_{\perp}\rho_{d,E}|\gg1$ in the inertial layer, do not contribute to Reynolds or Maxwell stresses. Here,  $\rho_d$ is the magnetic drift orbit width. EP nonlinearity enters implicitly in the curvature coupling term (CCT, second term on the left hand side of equation (\ref{vorticityequation})) in the ideal region via nonlinear EP response. The nonadiabatic EP response to ZF $\delta H^{NL}_Z$, is derived from the nonlinear gyrokinetic equation \cite{EFriemanPoF1982}:
\begin{eqnarray}
\left(-i\omega+v_{\parallel}\partial_l+i\omega_d\right)\delta H&=&-i\frac{e_s}{m}QF_0J_k\delta L_k\nonumber\\
&&-\frac{c}{B_0}\Lambda_kJ_{k'}\delta L_{k'}\delta H_{k''}\label{NLgyrokinetic}.
\end{eqnarray}
Here, $QF_0=(\omega\partial_E-\omega_*)F_0$ with $E=v^2/2$, $\omega_*F_0=\mathbf{k}\cdot\mathbf{b}\times\nabla F_0/\Omega$, $\omega_d=(v^2_{\perp}+2
v^2_{\parallel})/(2 \Omega R_0)\left(k_r\sin\theta+k_{\theta}\cos\theta\right)$,  $l$ is the length along the equilibrium magnetic field line, $J_k=J_0(k_{\perp}\rho)$ with $J_0$ being the Bessel function accounting for finite Larmor radius effects, $\langle\cdots\rangle$ indicates velocity space integration, $\delta L=\delta\phi-v_{\parallel}\delta A_{\parallel}/c$; and other notations are standard.

Linear EP response to TAE can be derived by transforming into drift orbit center coordinates. Assuming well circulating EPs for simplicity,
taking $\delta H^L_0=e^{i\lambda_{d0}}\delta H^L_{d0}$, with $\lambda_{d0}=\hat{\lambda}_{d0}\sin(\theta-\theta_0)=k_{\perp,0}\hat{\rho}_d\sin(\theta-\theta_0)$, $k_{\perp,0}=\sqrt{k^2_{\theta}+k^2_{0,r}}$, $\theta_0=\tan^{-1}(k_{0,r}/k_{\theta})$, $\hat{\rho_d}=qR_0\hat{v}_d/v_{\parallel}$, $\hat{v}_d=(v^2_{\perp}+2
v^2_{\parallel})/(2 \Omega R_0)$ and noting $\exp(ia\cos\theta)=\sum_l J_l(a)\exp(il\theta)$, we then have
\begin{eqnarray}
\delta H^L_0&=&-\frac{e}{m}Q_0F_0e^{i\lambda_{d0}}J_0(\gamma_0)\delta L_0\nonumber\\
&&\times\sum_l\frac{J_l(\hat{\lambda}_{d0})e^{il(\theta-\theta_0)}}{\omega_0-k_{\parallel,0}v_{\parallel}-l\omega_{tr}}.
\end{eqnarray}
Here, $e^{i\lambda_{d,0}}$ is the generator of coordinate transformation from drift orbit center to particle gyro center, $\omega_{tr}\equiv v_{\parallel}/(qR_0)$ is the transit frequency and $J_0(\gamma_0)=J_0(k_{\perp,0}\rho_L)$.  EP response to $\delta\phi^*_{0}$ can be derived similarly
\begin{eqnarray}
\delta H^L_{0^*}&=&-\frac{e}{m}Q_{0^*}F_0e^{i\lambda_{d0^*}}J_0(\gamma_{0^*})\delta L_{0^*}\nonumber\\
&&\times\sum_l\frac{(-1)^lJ_l(\hat{\lambda}_{d0^*})e^{il(\theta+\theta_{0^*})}}{\omega_{0^*}-k_{\parallel,0^*}v_{\parallel}-l\omega_{tr}}.
\end{eqnarray}
Here, $\lambda_{d0^*}=\hat{\lambda}_{d0^*}\sin(\theta+\theta_{0^*})=k_{\perp,0^*}\hat{\rho}_d\sin(\theta+\theta_{0^*})$, with $k_{\perp,0^*}=\sqrt{k^2_{\theta}+k^2_{0^*,r}}$ and $\theta_{0^*}=\tan^{-1}(k_{0^*,r}/k_{\theta})$.

Taking $\delta H^{NL}_Z=e^{i\lambda_{dZ}}\delta H^{NL}_{dZ}$ with $\lambda_{dZ}=\hat{\lambda}_{dZ}\cos\theta=k_Z\hat{\rho}_d\cos\theta$, we   have:
\begin{eqnarray}
\left(\partial_t+\omega_{tr}\partial_{\theta}\right)\delta H^{NL}_{dZ}=-\frac{c}{B}e^{-i\lambda_{dZ}}\Lambda_Z J_0(\gamma_{k'})\delta L_{k'}\delta H_{k''}.
\end{eqnarray}
Separating $\delta H^{NL}_{dZ}=\overline{\delta H^{NL}_{dZ}}+\widetilde{\delta H^{NL}_{dZ}}$, with $\overline{(\cdots)}$ and $\widetilde{(\cdots)}$ denoting surface averaged and  poloidally varying components, respectively; and noting $|\widetilde{\delta H^{NL}_{dZ}}/\overline{\delta H^{NL}_{dZ}}|\sim |\omega_Z/\omega_{tr}|\ll1$, we then obtain
\begin{eqnarray}
\partial_t\overline{\delta H^{NL}_{dZ}}&=&-\frac{c}{B_0}\overline{e^{-i\lambda_{dZ}}\Lambda_Z J_0(\gamma_{k'})\delta L_{k'}\delta H_{k''}},\label{NLdc}\\
\omega_{tr}\partial_{\theta}\widetilde{\delta H^{NL}_{dZ}}&=&-\frac{c}{B_0}\left[e^{-i\lambda_{dZ}}\Lambda_Z J_0(\gamma_{k'})\delta L_{k'}\delta H_{k''}\right]_{AC}\label{NLac}.
\end{eqnarray}
Here, the subscript ``AC" denotes $m\neq0$ component, and $(\cdots)_{AC}=\widetilde{(\cdots)}$.

Nonlinear EP response enters vorticity equation via surface averaged CCT contribution in the ideal region. Noting that $\omega_{dZ}=\omega_{tr}\partial_{\theta}\lambda_{dZ}$,   we  have
\begin{eqnarray}
\mbox{CCT}&=&\left\langle\overline{\frac{e}{\omega}J_0(\gamma_Z)\omega_d\delta H^{NL}}\right\rangle\nonumber\\
&=&-\frac{i}{2\pi}\frac{e}{\omega}\left\langle J_Z\int d\theta e^{i\lambda_{dZ}}\omega_{tr}\partial_{\theta}\widetilde{\delta H^{NL}_{dZ}}\right\rangle.\label{CCT1}
\end{eqnarray}
Here, $J_Z=J_0(k_Z\rho)$. It is readily obtained from equation (\ref{CCT1}) that, despite $|\widetilde{\delta H^{NL}_{dZ}}/\overline{\delta H^{NL}_{dZ}}|\ll1$, the contribution of EPs to CCT in the vorticity equation for the ZFZS
comes only from $\widetilde{\delta H^{NL}_{dZ}}$. Meanwhile, the flux surface averaged response, $\overline{\delta H_{dZ}^{NL}}$, would dominate the EP nonlinear wave-particle response in the TAE vorticity equation \cite{LChenRMP2016,FZoncaVarenna2000}. This is not the subject of the present work and will be treated elsewhere. Substituting equation (\ref{NLac}) into equation (\ref{CCT1}), and noting  that $\overline{A \widetilde{B}}=\overline{\widetilde{A}B}$ and $\widetilde{e^{i\lambda_{dZ}}}=e^{i\lambda_{dZ}}-J_0(\hat{\lambda}_{dZ})$, we then have
\begin{eqnarray}
\mbox{CCT}&=&\frac{i}{2\pi}\frac{c}{B_0}\frac{e}{\omega}\left\langle J_Z\left[\underbrace{\int d\theta\Lambda_ZJ_0(\gamma_{k'})\delta L_{k'}\delta H_{k''}}_{\mathscr{A}}\right.\right.\nonumber\\
&-&\left.\left.J_0(\hat{\lambda}_{dZ})\underbrace{\int d\theta e^{-i\lambda_{dZ}}\Lambda_ZJ_0(\gamma_{k'})\delta L_{k'}\delta H_{k''}}_{\mathscr{B}}\right]\right\rangle.\nonumber
\end{eqnarray}
$\mathscr{A}$ and $\mathscr{B}$ terms will be treated separately.

Using linear EP responses in the nonlinear terms (i.e., the linear expression for $\delta H_{k''}$ in the nonlinear term), ignoring the weak tunneling coupling between two poloidal harmonics located at different radial positions, and noting that $Q_{0^*}\simeq -Q_0$ due to $|\omega_{*,E}|\gg|\omega_0|$, we then have
\begin{eqnarray}
\mathscr{A}&=&-\hat{H}\int d\theta\left[J_0(\gamma_{0^*})\delta L_{0^*}\delta H_0-J_0(\gamma_0)\delta L_0\delta H_{0^*}\right]\nonumber\\
&=&2\pi \hat{H}\frac{e}{m}J_0(\gamma_0)J_0(\gamma_{0^*})\hat{A}_0\hat{A}_{0^*}\nonumber\\
&&\times\sum_m|\Phi_0|^2
\left(1-\frac{k_{\parallel}v_{\parallel}}{\omega}\right)_0\left(1-\frac{k_{\parallel}v_{\parallel}}{\omega}\right)_{0^*} Q_0F_0 \nonumber\\
&&\times\sum_l\left[\frac{J^2_l(\hat{\lambda}_{d0})}{\omega_0-k_{\parallel,0}v_{\parallel}-l\omega_{tr}}+\frac{J^2_l(\hat{\lambda}_{d0^*})}{\omega_{0^*}-k_{\parallel,0^*}v_{\parallel}-l\omega_{tr}}\right].\nonumber\\
\label{eq:A1}
\end{eqnarray}
Here, $\hat{H}=k_{\theta}(k_{r,0}+k_{r,0^*})$. In deriving equation (\ref{eq:A1}), we have applied ideal MHD condition ($\delta\phi^L\simeq\delta\psi^L$) for  TAEs to simplify $\delta L_0$ and $\delta L_{0^*}$ \cite{ZQiuEPL2013}.

Assuming that dominant contribution comes from resonant EPs,  we then have
\begin{eqnarray}
\mathscr{A}&=&2i\pi^2\hat{H}\frac{e}{m}J_0(\gamma_0)J_0(\gamma_{0^*})Q_0F_0\frac{\omega^2_{tr}}{\omega^2_0}\nonumber\\
&&\times\sum_l l^2\left(J^2_l(\hat{\lambda}_{d0})-J^2_l(\hat{\lambda}_{d0^*})\right)\delta (\omega_0-k_{\parallel,0}v_{\parallel}-l\omega_{tr})\nonumber\\
&&\times|\hat{A}_0|^2\sum_m|\Phi_0|^2.\label{eq:A}
\end{eqnarray}
In deriving equation (\ref{eq:A}), the resonance condition is applied to simplify $\delta L_k$ (i.e., $\omega-k_{\parallel,0}v_{\parallel}=l\omega_{tr}$). The contribution to the nonlinear term, comes from the finite orbit width  (FOW) effects   induced $k_{\perp}$-spectrum asymmetry, with an interesting analogue to the well-known polarization nonlinearity induced by the finite Larmor radius effect \cite{AHasegawaPoF1978}. One would then expect, comparing to the well-circulating EPs assumed here, trapped EPs may enhance the nonlinear couplings even stronger  due to their large bounce orbits. This will be discussed  in a future publication.

$\mathscr{B}$ can be derived similarly. Substituting linear EP response ($\delta H_{k''}$) into $\mathscr{B}$, and noting $k_Z=k_{r,0}+k_{r,0^*}$,  we obtain:
\begin{eqnarray}
\mathscr{B}&=&\hat{H}\frac{e}{m}J_0(\gamma_0)J_0(\gamma_{0^*})\int d\theta e^{-i\lambda_{dZ}}\delta L_0\delta L_{0^*}Q_0F_0\nonumber\\
&&\times\left[e^{-i\lambda_{d0}}\sum_l\frac{J_l(k_{\perp,0}\hat{\rho}_d)e^{il(\theta-\theta_0)}}{\omega_0-k_{\parallel,0}v_{\parallel}-l\omega_{tr}}\right.\nonumber\\
&&\hspace*{1.5em}\left.+e^{i\lambda_{d0^*}}\sum_l\frac{(-1)^lJ_l(k_{\perp,0^*}\hat{\rho}_d)e^{il(\theta+\theta_{0^*})}}{\omega_{0^*}-k_{\parallel,0^*}v_{\parallel}-l\omega_{tr}}\right]\nonumber\\
&=&0.\nonumber
\end{eqnarray}

Assuming $|k_{\perp}\rho_{d,E}|\ll1$, and keeping only $l=\pm1$ transit resonances, we then have
\begin{eqnarray}
\mbox{CCT}&=&\frac{i}{4}\pi\frac{c}{B_0}\frac{e^2}{m}\frac{n_{0E}}{\omega_Z}\frac{k_{\theta}}{\omega^2_0}\hat{G}\frac{\partial^2}{\partial r^2}\hat{F}|\hat{A}_0|^2\sum_m|\Phi_0|^2.\nonumber
\end{eqnarray}
Here, $\hat{F}\equiv i(\hat{k}_{r,0}-\hat{k}_{r,0^*})+\partial_r\ln\Phi_0-\partial_r\ln\Phi_{0^*}$, with $\hat{k}_{r,0}-\hat{k}_{r,0^*}$ from radial envelope modulation and $\partial_r\ln\Phi_0-\partial_r\ln\Phi_{0^*}$ related with fine radial structures of TAE  \cite{ZQiuNF2016}. $\hat{G}$ comes from resonant EP, and is defined as
\begin{eqnarray}
\hat{G}&\equiv&\left\langle \omega_{*,E}\hat{v}^2_d(F_{0E}/n_{0E})\right.\nonumber\\
&&\left.\times\left(\delta(\omega_0-k_{\parallel}v_{\parallel}-\omega_{tr})+\delta(\omega_0-k_{\parallel}v_{\parallel}+\omega_{tr})\right)\right\rangle\nonumber.
\end{eqnarray}
In the expression of $\hat{G}$, the FLR effects are ignored in consistency with the $k_{\perp}\rho_{d,E}\ll1$ assumption.

Thermal plasma contribution to nonlinearity  comes from Reynolds (RS) and Maxwell (MX) stresses in the inertial layer. We have, following Ref. \citenum{LChenPRL2012} \footnote{Note that in Ref. \citenum{LChenPRL2012}, the fine structure of ZF is not considered, such that there is no $\partial_r\ln\Phi_0-\partial_r\ln\Phi_{0^*}$ term.}:
\begin{eqnarray}
\mbox{RS+ MX}&=&-\frac{1}{2}\frac{c}{B_0}\frac{n_0e^2}{T_i}k_{\theta}\rho^2_i\frac{1}{\omega_Z}\left(1-\frac{k^2_{\parallel}v^2_A}{\omega^2}\right)\nonumber\\
&&\times\frac{\partial^2}{\partial r^2}\hat{F}|\hat{A}_0|^2\sum_m|\Phi_0|^2.
\end{eqnarray}

Noting that the EP induced nonlinearity dominates over Reynolds and Maxwell stresses by order $O(n_{0E}\hat{\omega}_{*E}q^2/(n_0\omega_0\epsilon))$, the nonlinear vorticity equation for ZF then becomes
\begin{eqnarray}
\noindent\omega_Z\hat{\chi}_{iZ}\delta\phi_Z=i\frac{\pi}{4}\frac{c}{B_0}\frac{n_{0E}}{n_0}\frac{T_i}{T_E}\frac{k_{\theta}}{\rho^2_i\omega^2_0}\hat{G}\hat{F}|\hat{A}_0|^2\sum_m|\Phi_0|^2.\label{ZFVorticity}
\end{eqnarray}
Here, $\hat{\chi}_{iZ}\equiv\chi_{iZ}/(k^2_r\rho^2_i)\simeq 1.6q^2/\sqrt{\epsilon}$  with $\chi_{iZ}$ being the neoclassical polarization \cite{MRosenbluthPRL1998}, $\hat{\omega}_{*E}\equiv T_E \mathbf{k}\cdot\hat{\mathbf{b}}\times\nabla\ln F_{0E}/(m_i\Omega_i)$ and  $\epsilon\equiv r/R_0\ll1$ being the inverse aspect ratio \footnote{Note that in deriving equation (\ref{ZFVorticity}), we assumed small EP drift orbit in the ideal region. The same equation can be obtained by assuming that EP response is dominated by the $l=\pm 1$ transit resonances, while no assumptions on EP drift orbit is needed.}.

For  $\Phi_0$ being purely real, the ZF  generation rate is dominated by the first term of $\hat{F}$ (i.e. $\hat{k}_{r,0}-\hat{k}_{r,0^*}$), which corresponds to radial envelope modulation. This is the typical case for fixed  shear Alfv\'en waves with a prescribed amplitude \cite{LChenPRL2012} (the nonlinear term in equation (\ref{ZFVorticity}) should be replaced by RS and MX, but the structure of the nonlinear term is not changed)  and/or drift waves \cite{LChenPoP2000}. On the other hand, for  the case of EP driven TAE discussed here, $\Phi_0$ is complex due to wave-particle interactions, and thus, the second term (i.e., $\partial_r\ln\Phi_0-\partial_r\ln\Phi_0^*$) is finite, and dominates. In this case, the generation rate is enhanced by $O(1/\hat{k}_{Z}\Delta_s)$, with $\Delta_s$ being the scale of the fine structure which is, typically,    distance between mode rational surfaces. The generated ZF, in addition to the usual ``meso"-scale,   also has a fine-scale  radial structure \cite{ZQiuNF2016}.

Keeping only the dominant term associated with TAE fine radial structure, we then have
\begin{eqnarray}
\partial_t\hat{\chi}_{iZ}\delta\phi_Z=i\frac{\pi}{2}\hat{K}\hat{G}\mathbb I{\rm m}(\partial_r\ln\Phi_0)|\hat{A}_0|^2\sum_m|\Phi_0|^2,\label{eq:ZFtemporal}
\end{eqnarray}
with $\hat{K}\equiv cn_ET_ik_{\theta}/(B_0n_0T_E\rho^2_i\omega^2_0)$  defined consistently with equation (\ref{ZFVorticity}) by direct inspection. For TAE with a finite   growth rate $\gamma_L$  due to   EP resonant drive, we then have, $\partial_t|_Z=2\gamma_{\scriptsize L}$. The generation of ZF discussed here is a typical forced driven process, consistent with  simulation results \cite{YTodoNF2010,ZWang2016}. This process   is different from that of modulational instability \cite{LChenPRL2012}, which, dubbed as ``secondary instability", becomes important as the pump wave reaches a certain amplitude to overcome the threshold condition for reinforcement by nonlinearity of its deviation from periodic behavior; while the forced driven process studied here, occurs while the pump wave is still in the linear growth stage. The forced driven process  is, thus, expected to have potentially significant consequences on TAE nonlinear dynamics.

Noting again $\partial_t|_Z=2\gamma_L$, the generated ZF can then be derived
\begin{eqnarray}
\delta\phi_Z=i\frac{\pi}{4}\frac{\hat{K}\hat{G}}{\gamma_{\scriptsize L}\hat{\chi}_{iZ}}\mathbb I{\rm m}(\partial_r\ln\Phi_0)|\hat{A}_0|^2\sum_m|\Phi_0|^2.\label{eq:ZF}
\end{eqnarray}
It is clear from equation (\ref{eq:ZF}) that ZF has both a meso-scale and a fine-scale radial structure, with the fine structure  $\Phi_Z$  related to $|\Phi_0|^2$. Taking
\begin{eqnarray}
\Phi_Z\equiv|\Phi_0|^2,
\end{eqnarray}
 the meso-scale structure of ZF is then
\begin{eqnarray}
\hat{A}_Z=i\frac{\pi}{4} \frac{\hat{K}\hat{G}}{\gamma_{\scriptsize L}\hat{\chi}_{iZ}} \mathbb I{\rm m}(\partial_r\ln\Phi_0)|\hat{A}_0|^2.
\end{eqnarray}

In conclusion, the nonlinear excitation of ZF by EP driven TAE is studied, and it is found that  EP contribution in the ideal region may dominates over the usual Reynolds and Maxwell stresses in the layer region. In addition to the secondary modulational process discussed in Ref. \citenum{LChenPRL2012}, ZF can also be excited by forced driven process in the linear growth stage of TAE \cite{YTodoNF2010,ZWang2016}. The growth rate of the forced driven ZF is twice of   TAE   growth rate, and the generated ZF has both the usual meso scale and  fine radial structure, due to the fact that AEs are typically weak or moderate ballooning \cite{ZQiuNF2016}. The nonlinear coupling effect between AEs  is of order $O(1/\hat{k}_Z\Delta_s)$ stronger comparing to envelope modulation, when  the anti-Hermitian response due to EP resonant drive (more generally, wave-particle interaction) and the corresponding complex fine radial TAE structure is properly taken into account. The mechanism for ZF drive here, is polarization induced by   resonant EP  nonlinearity.

We note that  the forced driven mechanism for ZF generation, discussed here, is very different from that of spontaneous excitation   via modulational instability. Modulational instability becomes important when the amplitude of the pump wave (TAE here) is large enough to overcome the threshold condition  due to, e.g., frequency mismatch \cite{LChenPRL2012} and/or dissipations. On the other hand,  the forced driven process, being essentially thresholdless, takes place in the initial linear growth stage of the pump wave ($|\omega_{B}|<|\gamma_L|$, with $\omega_{B}$ being the wave particle trapping frequency, proportional to the square root of mode amplitude), and may have significant consequences on the nonlinear dynamics of the pump TAE. First of all, with the  growth rate being twice  TAE linear growth rate,  EP resonance detuning by  ZF may compete with   phase space wave-particle nonlinearities. Second, the forced driven ZF may regulate the   saturation level of TAE. Equation (\ref{eq:ZFtemporal}) shows that, after the initial exponential growth and as TAE  saturates with $\gamma_{TAE}\rightarrow0$, the temporal evolution of ZF   becomes  algebraic, which can be suppressed by, e.g.,  collisional damping.  If the saturation level of TAE determined by forced driven ZF exceeds the threshold condition for modulational instability,  ZF and   TAE upper/lower sidebands    can   be generated   with   growth rate proportional to pump TAE amplitude  \cite{LChenPRL2012}.
Vice-versa, if the saturation level of TAE due to regulation by the forced driven ZF  is below  the modulational instability threshold,   the spontaneous excitation process can be  completely suppressed \footnote{Note that phase space wave-particle nonlinearities is not included in the present system yet \cite{LChenRMP2016,FZoncaNJP2015},  as anticipated in our comments following equation (\ref{CCT1}).}. To correctly understand the nonlinear dynamics of Alfv\'en waves, all these mechanisms, including nonlinear wave-particle interactions \cite{HBerkPoFB1990c,FZoncaNJP2015,LChenRMP2016}  and nonlinear mode-mode couplings \cite{LChenPRL2012,LChenRMP2016}, should be taken into account on the same footing. The formulation of  such general problem and the derivation of the governing nonlinear equations  will be reported in a future publication.

This work is supported by     US DoE GRANT,  the ITER-CN under Grants Nos.  2013GB104004  and   2013GB111004,
the National Science Foundation of China under grant Nos.  11575157  and 11235009,  Fundamental Research Fund for Chinese Central Universities under Grant No. 2016FZA3003 and   EUROfusion Consortium
under grant agreement No. 633053.

\end{document}